# Who farted?
## Hydrogen sulphide transport from Bardarbunga to Scandinavia

Håkan Grahn, Pontus von Schoenberg, Niklas Brännström

The Swedish Defence Research Agency, Umeå

## Abstract

On September 9 2014 several incidences of foul smell (rotten eggs) were reported on the coast of Norway (in particular in the vicinity of Molde) and then on September 10 in the interior parts of county Västerbotten, Sweden. One of the theories that were put forward was that the foul smell was due to degassing of the Bardarbunga volcano on Iceland. Using satellite images (GOME-1,-2) of the sulphur dioxide, $SO_2$, contents in the atmosphere surrounding Iceland to estimate flux of $SO_2$ from the volcano and an atmospheric transport model, PELLO, we vindicate this theory: we argue that the cause for the foul smell was hydrogen sulphide originating from Bardarbunga.

The model concentrations are also compared to $SO_2$ concentration measurements from Muonio, Finland.

## Introduction
On September 9 2014 the fire brigade in Molde, Norway, was called to investigate a case of foul smell akin to rotten eggs. When the fire brigade arrived they were unable to sense the smell but by then they had received many more reports of the same phenomenon. Norwegian meteorologists pointed to the active volcano Bardarbunga, Iceland, suggesting it was the source of the foul smell [NRK]. The day after similar incidences of foul smell, described in colourful language in the press, were reported from Storuman, county Västerbotten, Sweden [VK]. Swedish meteorologists were reported not to be as confident as their Norwegian counterparts pointing to other possible reasons like air pollution stemming from northern Europe [VK]. The same smell was then reported from county Troms, Norway, coupled with elevated $SO_2$ measurements in neighbouring county Finnmark, Norway [WSJ].

The Bardarbunga volcanic system had been active since the middle of August 2014 [IMO], and, as evident from satellite images of $SO_2$ [OMI], [GOME-2] in the atmosphere around Iceland, it had been degassing since the last few days of August 2014 (an effusive eruption with no release of volcanic ash into the atmosphere). Volcanic gasses have many constituents, with $H_2O$, $CO_2$ and $SO_2$ being the dominant ones, but there are other ones as well, notably $H_2S$ [Le Guern]. Given the descriptive smell that was reported we assume that is was caused by a sulphuric compound. In the present case the measurements we are relying on are conducted by the human nose: the olfative threshold (the threshold for noticing smell) for $SO_2$ lies in the range 1.175 $mg/m^3$ to 12.5 $mg/m^3$, while for $H_2S$ the range is 0.0007 $mg/m^3$ to 0.0140 $mg/m^3$ according to [Ruth1986]. In [Guidotti1994] the olfactive threshold for $H_2S$ is reported to be 0.0150 $mg/m^3$ together with a remark that it is highly variable.

In this letter we employ an atmospheric transport model to study whether these stated concentration thresholds could have been exceeded following the degassing of Bardarbunga. To complete such a study a number of ingredients are required. First we need an estimate of the flux of $SO_2$ and $H_2S$ from the volcano: that is we need a source term. Secondly we need an atmospheric transport model which ideally handles any atmospheric chemistry internally. In lieu of this we treat atmospheric chemistry as an exogenous factor, thus as a third component in the study we need some input on the sulphur chemistry taking place in the atmosphere. In addition to this we need an idea of the chemistry taking place as the gasses are released from the volcano. Volcanic degassing is complex, as is atmospheric chemistry and atmospheric transport modelling. We will make a number of crude simplifications in this study, and they will be stated in the next section, but despite this we will argue that the olfactive threshold for $H_2S$ was exceeded in Storuman, Sweden.

The letter is organised as follows: in the next section we describe the source term estimate, the atmospheric transport model, and our simplifications of the atmospheric and volcanic chemistry. Following that we present our results followed by a discussion. The letter is rounded off with some concluding remarks.

## Method

### PELLO – A Lagrangian random displacement model

PELLO is a dispersion model developed at the Swedish Defence Research Institute, FOI, [Lindqvist 1999]. The model is in use by the Swedish Radiation Safety Authority, SSM. It is a Lagrangian random displacement model where model particles represent the dispersed material, gas or aerosol. The model particles are tracked on their way through the atmosphere. In each time step every model particle is transported by the wind field and a random displacement movement is added to represent the current turbulence. This way the dilution of the cloud is described. The model takes into account dry deposition of gases, and wet and dry deposition for particles, along the path. To summarize the result box counting is used to calculate concentration and deposition. More details on the model will be published elsewhere.

### Volcanic and atmospheric chemistry, simplifications

In [Aiuppa et al 2007] the relative amounts of different chemical compounds in volcanic gases emitted from Mount Etna and Vulcano Island, both volcanoes situated in Sicily, were measured. Measurements were taken at the vents (or to be precise, 0.1 km downwind the vents) as the gases were released from the volcano and then again at 10 km downwind from the vents. They showed that the relative molar ratio $SO_2/H_2S$ was 60 at the vents. They could not detect any systematic change in this molar ratio with plume aging (the age of the measured plumes were typically a few minutes old up to a few hours). See also [Aiuppa et al 2008].

*Assumption:* We assume that the volcanic gases from Bardarbunga has a molar ratio $SO_2/H_2S$ equal to 60. We further assume that this ration remains constant as the plume ages, i.e. we assume that the molar ratio remains constant for all times.

As we will be working with concentrations of $SO_2$ and $H_2S$ (measured in mass/volume air) it is more natural to work with mass ratios than molar ratios. A molar ratio of 60 corresponds to a mass ratio of 113.

Currently the dispersion model PELLO treats all model particles as inert. In the present study regarding long range transport of $SO_2$, this is a drawback as $SO_2$ will interact with the atmospheric gases and form $SO_4$ -aerosols. In the literature there is, however, disagreement concerning the rate at which this happens. For the Laki 1783 eruption, a historical eruption on Iceland, [Grattan et al 2003] states that no $SO_2$ was converted to $SO_4$-aerosols during the transport from Iceland to mainland Europe, while [Oman et al 2006] states that 70% was converted (based on simulations) and [Stevenson et al 2006] puts the figure at 30% (also based on simulations).

*Assumption:* We will assume that $SO_2$ is an inert gas during the simulations and accordingly the results presented in the Results section will be in line with this assumption. However, we will return to this issue in the Discussion.

### Source reconstruction

An essential input to the atmospheric transport model is the flux of gases from the volcano. In this case we are interested in the flux of $SO_2$. The amount of $SO_2$ in the atmosphere is routinely monitored by satellites, operating for example in the ultraviolet-visible band: OMI/Aura (see eg. [OMI])) and GOME-2/MetOp-A (see eg. [GOME-2]). For our purposes satellite observations come with a drawback, they typically yield integrated amounts of previously dispersed gas. There are a number of methods available, reviewed in [Theys et al], for converting satellite measured masses of $SO_2$ into $SO_2$ fluxes (emission rates). The methods range from simpler ones [Lopez et al] estimating the flux by M*v/L (where M is the total mass as measured by the satellite, v is the wind speed, and L is the length of the plume in the direction of transport) to more involved ones relying on solving the inverse atmospheric dispersion problem [Stohl et al] incorporating effects of $SO_2$ atmospheric chemistry.

We have employed a method which may be considered as a simple prelude to solving the full inverse problem as described by [Stohl et al]. We impose the simplifying assumption that the only depletion of $SO_2$ takes place by dry deposition, that is, we ignore the effect of chemical reactions (such as $SO_2$ forming $H_2SO_4$). Then we assume that for every 24 hour period the emission rate is constant. We are interested in estimating the source term for the period 29 August 2014 to 9 September 2014, thus estimating the source term amounts to assigning constant release rates to 12 day-long emissions. Each satellite image covers the lat-long box 20°W-20°E, 60°N-70°N, and consequently this is the area for which the total mass of $SO_2$ is given. This is now set up as a classic inverse problem: given the satellite observations determine the daily emission rates from the volcano (we assume that the volcano is the only $SO_2$ source in the area covered). We solve this inverse problem heuristically by scaling the emission rates on each day until the total mass of $SO_2$ in the box matches that of the satellite measurements. From a practical point of view this is done by letting our dispersion model PELLO simulate the transport from a continuous unit source on each day (by a unit source we denote a source that releases 1 kton $SO_2$ each day starting at 00:00 and ending at 24:00), then we integrate the concentration of $SO_2$ in the lat-long box 20°W-20°E, 60°N-70°N to obtain the total mass of $SO_2$ that this unit source has given rise to (within the lat-long box). Then, for every day, we superposition the total mass of $SO_2$ that these unit sources have given risen to ($SO_2$ released on a given day may linger for several days with the given lat-long box). By scaling the source strengths of the unit sources we try to match the total mass of $SO_2$ in the lat-long box to those observed by the satellites. If this procedure is successful, as it is in the present case, it yields a feasible solution to the inverse problem. It is however not necessarily the best solution to the inverse problem (we do not claim that the solution is optimal in any sense nor the most probable one). Applying the method outlined above we estimate the daily flux of $SO_2$ to be that given in Figure 1.

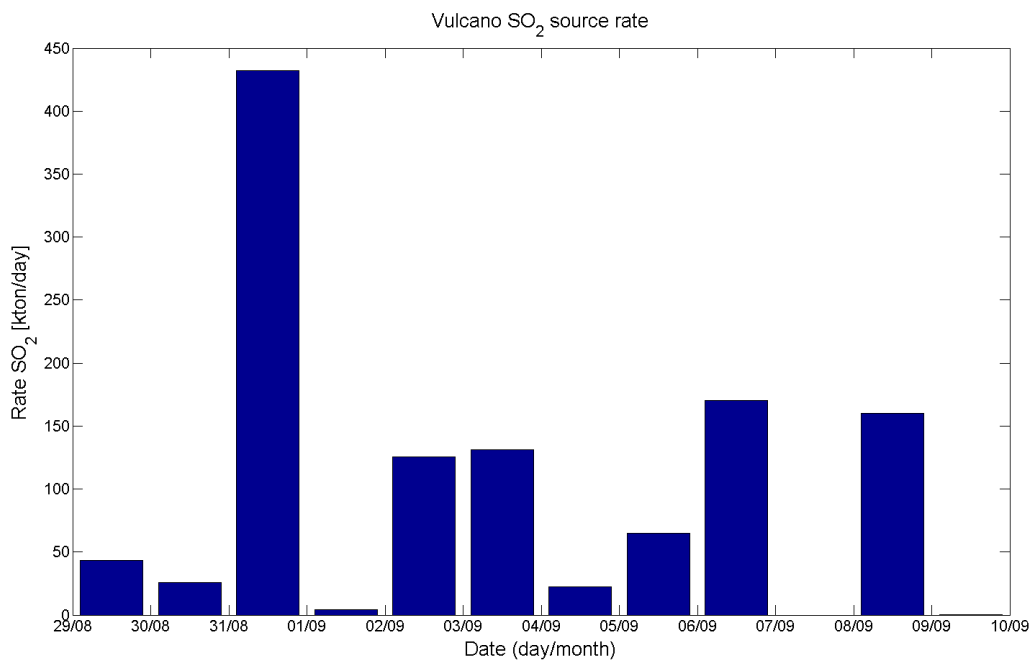

Figure 1: The estimated $SO_2$- source term, release rates is kton/day.

The simulated total mass of $SO_2$ in the lat-long box closely matches the satellite measured total mass of $SO_2$, see Figure 2.

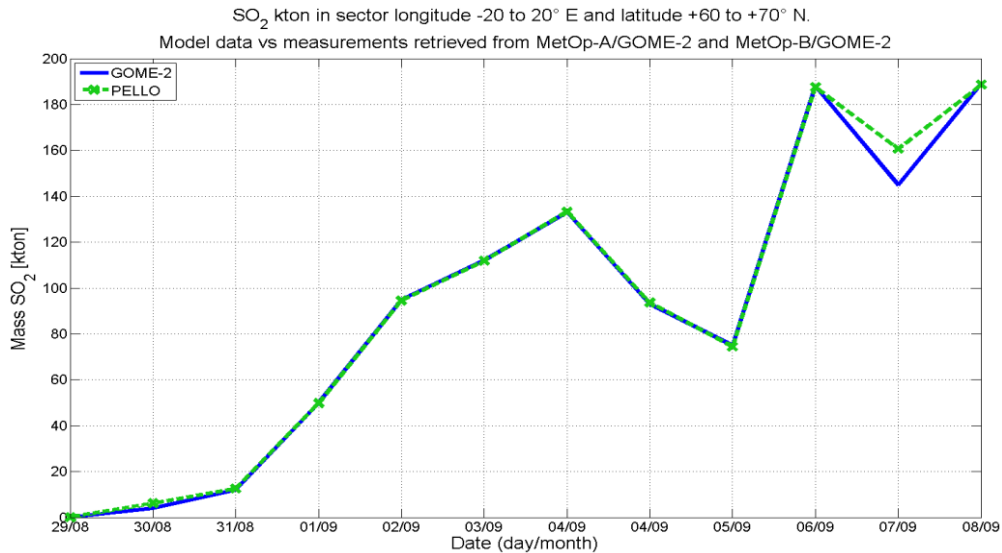

Figure 2: Total mass $SO_2$ in the atmosphere in the lat-long box 20°W-20°E, 60°N-70°N the blue solid line being the satellite measurements [GOME-2], and the red dashed line being the simulated total mass in the same box using the estimated source presented in Figure 1. Measurement points are at 00:00 each day. In a normal situation the total mass fluctuates between 0 and 5 kton. The overestimation of simulated total mass at 00:00 September 8 is due to the source term on September 6.

## Results

Using the source given in Figure 1 and letting the dispersion model PELLO simulate the dispersion of $SO_2$ on the Northern Hemisphere we can extract spatio-temporal concentrations in given locations. Foul smell was reported in Molde, Norway, and in Storuman, Sweden, and in addition to this $SO_2$ measurements from Muonio, Finland, and Karpdalen, Norway, was reported in [WSJ]. We therefore report the simulated $SO_2$ concentrations, see Figure 3, and corresponding $H_2S$ concentrations, see Figure 4, in Storuman, Sweden, and Viksjøfjell (close to Karpdalen) and Molde, both Norway, and Muonio, Finland. PELLO store concentration data at each simulated hour and we plot that concentration in the figures, called 1 hour model result, together with a 12 hour moving mean value.

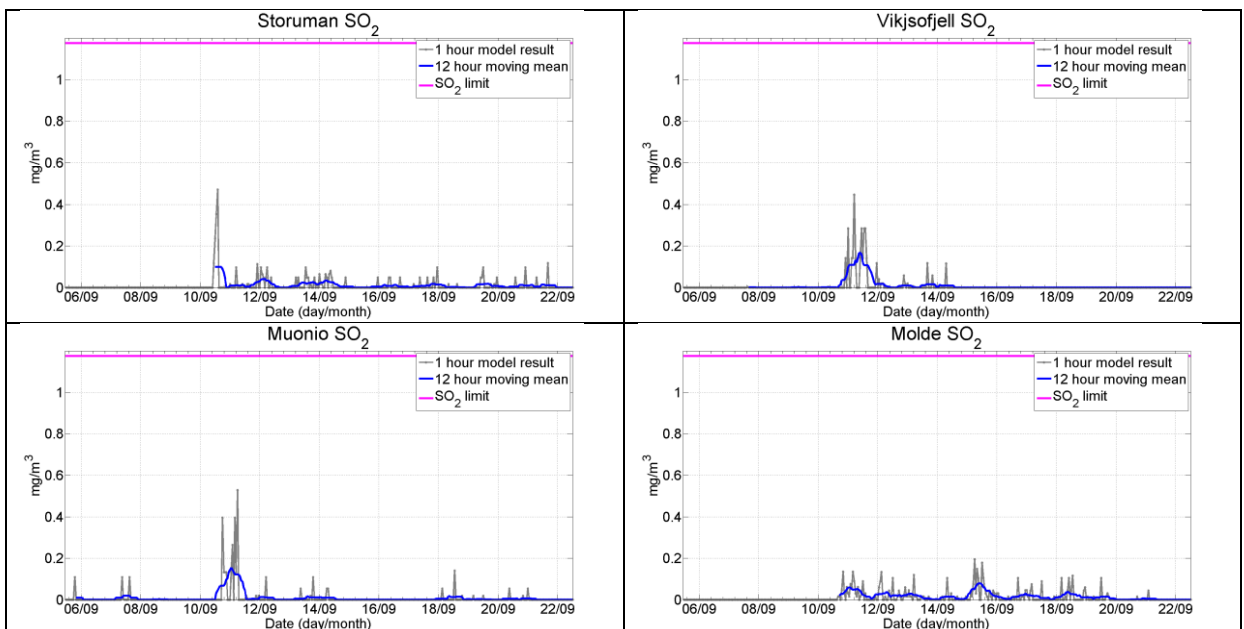

Figure 3: Model results: $SO_2$-concentration in $mg/m^3$ for Storuman, Viksjøfjell, Muonio and Molde. The olfactive threshold 1.175 $mg/m^3$ is represented by a horizontal purple line. Note that the scale of the concentration axis is linear.

Using the assumptions stated in the Methods section that the mass ratio for $SO_2/H_2S$ is 113 the corresponding model concentrations for $H_2S$ is given in Figure 4.

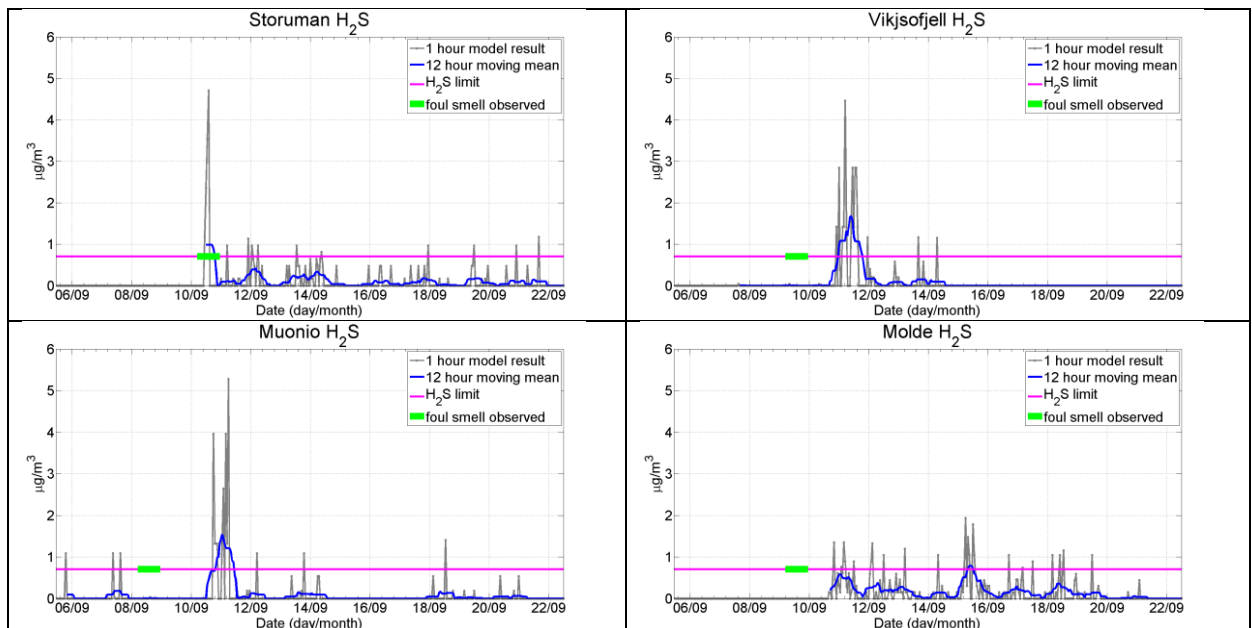

Figure 4: Model results: $H_2S$ -concentration in $mg/m^3$ for Storuman, Viksjøfjell, Muonio and Molde. The olfactive threshold 0.7 $\mu g/m^3$ is represented by a horizontal purple line. Note that the scale of the concentration axis is linear.

## Discussion

Let us consider the $SO_2$ and $H_2S$ concentration profiles in Figures 3 and 4 and compare these with the olfactive threshold for $SO_2$ and $H_2S$: respectively 1.175 $mg/m^3$ to 12.5 $mg/m^3$, and 0.0007 $mg/m^3$ to 0.015 $mg/m^3$. Based on our simulation results we conclude that it is unlikely that $SO_2$ is to blame for the foul smell (although, in Storuman the peak $SO_2$ concentration is within one order of magnitude of the olfactive threshold), while it is likely that $H_2S$ exceeded the olfactive threshold.

### Nose vs. measurements

Although we set out to test the hypothesis that $H_2S$ originating from Bardarbunga was the source of the foul *smell* reported in local media on September 8, 9, and 10 2014 it is of course of interest to compare the model results with measurements made by scientific sensors rather than noses. In [WSJ] $SO_2$ concentrations for 8 September where reported in Muonio, Finland at 0.187 $mg/m^3$ and 9 September at 0.150 $mg/m^3$ at Karpdalen, Norway. Our model results under-predict the measurement in Karpdalen by one order of magnitude, but the timing of our concentration peak (1 hour model results) is in line with the reported observation, see Figure 5.

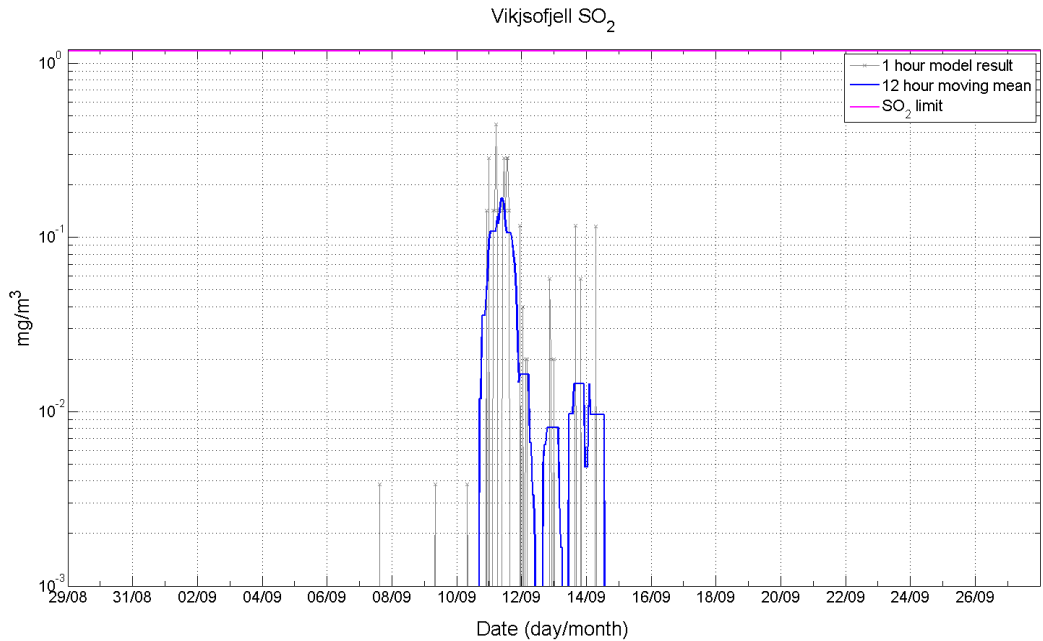

**Figure 5** Model results: $SO_2$-concentration in $mg/m^3$ for Viksjøfjell (close to Karpdalen), Norway. The olfactive threshold 1.175 $mg/m^3$ is represented by a horizontal purple line. Note that the scale of the concentration axis is logarithmic and not linear.

The source of the Finnish measurements referred to in [WSJ] is [Ilmanlaatu], where hourly measurements (but still unverified) for all of September 2014 are presented which makes it a good data set to bench mark against. In Figure 6 we have plotted our model concentrations against measurements retrieved, by visual inspection, from [Ilmanlaatu]. Comparing our model results to the measurements we note that

1. we miss the peak 16-18 Sept (we disregard the peaks after 22 September as we have not tried to estimate the source term that would explain those peaks)
2. Our 1 hour model results are of the same order of magnitude as the measurement (these are 1 hour measurements)
3. Our 12 hours moving mean is within an order of magnitude from the measurements (remember these are 1 hour measurements)
4. For the first two peaks our model results are lagging in time by approximately 12-24 hours and for the two remaining peaks (disregarding the peak 12-18 Sept which our model fails to explain) our model results precede the measurements by approximately the same time discrepancy.

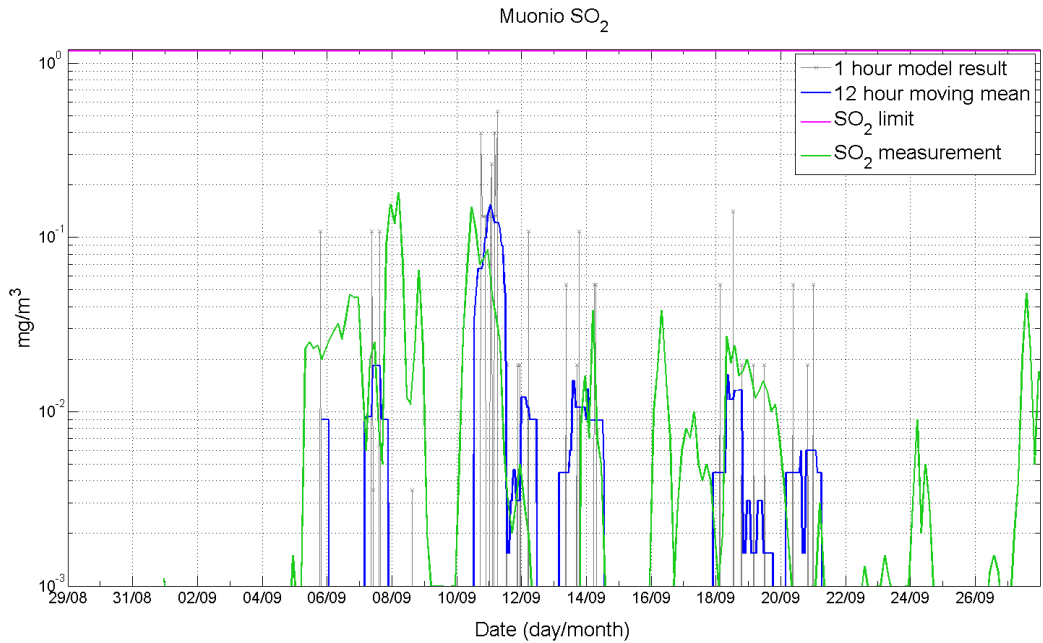

Figure 6 Model results: $SO_2$-concentration in $mg/m^3$ for Muonio. The olfactive threshold 1.175 $mg/m^3$ is represented by a horizontal purple line. Note that the scale of the concentration axis is logarithmic and not linear.

The timing of the model concentration peak in Storuman corresponds well with the reported foul smell [VK].

By private communication with Ireland's Environmental Protection Agency we have also gained access to unverified $SO_2$ measurements from Irish weather stations. Ireland was exposed to elevated $SO_2$ concentrations on September 6 (sharp peaks). In this case our model is not capturing the situation all too well; in the first few days of September the weather situation brought the $SO_2$ cloud from Iceland towards Ireland and the data reveals sharp peaks of $SO_2$ at the Irish stations. In our simulation the cloud of $SO_2$ just brushes the western coast of Ireland and at a delay of 24 hours (the simulated concentrations under-predict the measured ones by a factor 6 in western Ireland, and are several orders of magnitude off in eastern Ireland).

### Errors and uncertainties

The dispersion model results that we have presented are associated with a number of uncertainties and errors. To begin with the source estimate (based on associating a flux of $SO_2$ from the volcano to the total amount of $SO_2$ in the atmosphere surrounding Iceland) is crude: we only have satellite images every 24 hours in the lat-long box 20°W-20°E, 60°N-70°N. In the W-E direction this is sufficient, but in the N-S direction we fear that the box may be too small allowing for $SO_2$ to be released from the volcano and transported out of the box within 24 hours rendering a too low source term in our estimate (see e.g. Figure 6 for an unfavourable weather situation for our source reconstruction method). Also, the 24 hour resolution means that our source term is a 24 hour average of the source (we assume the source term is piecewise constant every 24 hours). Since we are interested in concentration peaks (comparing against reported maximum measurements or exceedance of the olfactive thresholds) this can be limiting, especially if the volcanic degassing process is highly volatile in time (the simulated concentration peaks may be significantly reduced and shifted in time by up to 24 hours).

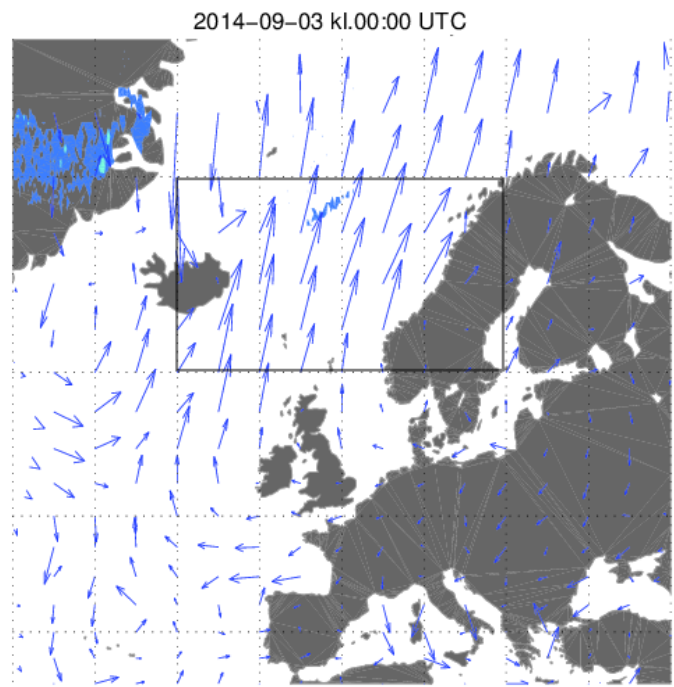

Figure 7 A plot of the wind field on Sept 3 2014. Any $SO_2$ released from Iceland would quite soon be transported out of the lat-long box 60°N-70°N, 20°W-20°E (black box) rendering too little $SO_2$ in the satellite image which in turn is used to reconstruct the strength of the source.

Secondly, we have assumed that the $SO_2/H_2S$ ratio remains constant throughout the simulation using $SO_2$ as a proxy for $H_2S$. The literature [Aiuppa et al 2012] indicates that this assumption is fair in the vicinity of the volcano (or rather for the two Italian volcanoes that they studied), but for long range transport we have no evidence that this assumption is valid. Thirdly, we have used $SO_2$ as a proxy for $H_2S$, but we have not taken into account the atmospheric chemistry that affects the amount of $SO_2$ in the plume; in particular the conversion of $SO_2$ to $SO_4$-aerosols. We treated $SO_2$ as an inert gas. The simulated concentration of $SO_2$ may be post-processed to take this effect into account. As stated previously there is no consensus regarding the conversion rate of $SO_2$ to $SO_4$-aerosols, but even if we use the conservative estimate, [Oman et el 2006], that 70% of the $SO_2$ has been converted to $SO_4$-aerosols when the plume reaches Scandinavia it seems likely that $H_2S$ was the cause for the foul smell in Storuman, Sweden.

As a next step it would be interesting to upgrade the model with aerosol chemistry and physics to better represent the physical processes that $SO_2$ undergoes while being transported.

## Conclusions

Based on satellite measurements of $SO_2$ in the atmosphere in the lat-long box 20°W-20°E, 60°N-70°N and the atmospheric dispersion model PELLO we estimated the average daily flux of $SO_2$ from the Bardarbunga volcano (presented in Figure 1). The source estimate is likely to be an underestimate of the source.

Using the estimated source and simulations using PELLO (atmospheric dispersion model) we conclude that the smell threshold for $H_2S$ was exceeded in Storuman, Sweden, and Viksjøfjell (close to Karpdalen) and Molde, both Norway, and Muonio, Finland. Furthermore we conclude that it is unlikely that $SO_2$ is to blame for the foul smell (although, in Storuman the peak $SO_2$ concentration is within an order of magnitude of the olfactive threshold).

Comparison with measurements in Muonio our model concentrations are in acceptable agreement considering the errors and uncertainties associated with the model.

## References

[NRK] Meteorolog: - Gasslukten på norskekysten kan stamme fra Bárðarbunga-vulkanen www.nrk.no/mr/melder -om-gasslukt-over-stort-omrad-1.11923138 (in Norwegian), retrieved 20 October 2014. Norsk ringkringkasting

[VK] Fisdimma över västerbotten www.vk.se/1273317/fisdimma-over-vasterbotten (in Swedish), retrieved 20 October 2014, Västerbottens-Kuriren

[WSJ] Iceland Volcanic Eruption Sending Toxic Gases Throughout Region, online.wsj.com/articles/iceland-volcanic-eruption-sending-gases-throughout-region-1410446383, retrieved 20 October 2014, Wall Street Journal Europe

[IMO] Icelandic Met Office, www.vedur.is, 2015

[GOME-2] Forschungszentrum der Bundesrepublik Deutschland für Luft- und Raumfahrt. Satellite images from GOME-2 (MetOp A&B) atmos.eoc.dlr.de/gome2/images/Bardarbunga_2014.gif, retrieved 20 Oct 2014.

[OMI] NASA. Satellite images from OMI/Aura http://so2.gsfc.nasa.gov/pix/daily/0914/iceland_so2lf_5k_ts_plot.html retrieved 29 Sept 2014.

[La Guern] F. Le Guern, Field measurements of volcanic gases, Abstracts of Papers presented to the Workshop on Remote Sensing of Volcanic Gasses: Current Status and Future Directions. Ed. Thomas R. McGetchin and T. B. McCord, published by the Lunar and Planetary Inst, 1979.

[Ruth1986] J. H. Ruth, Odor Thresholds and Irritation Levels of Several Chemical Substances: A Review, Am. Ind. Hyg. Assoc. J. (47), pp 142-151, 1986

[Guidotti1994] T. L. Guidotti, Occupational exposure to hydrogen sulfide in the sour gas industry: some unresolved issues, Int. Arch. Occup. Environ. Health (66), pp153-160, 1994

[Lindqvist 1999] En stokastisk partikelmodel I ett ickemetriskt koordinatsystem, *FOI Report, FOI-R—99-01086-862-SE*. FOI. 1999

[Aiuppa et al 2007] A. Aiuppa, A. Franco, R. von Glasow, A. G. Allen, W. D'Allessandro, T. A. Mather, D. M. Pyle, and M. Valenza, The tropospheric processing of acidic gases and hydrogen sulphide in volcanic gas plumes as inferred from field and model investigations, Atmos. Chem. Phys., 7, pp 1441-1450, 2007

[Aiuppa et al 2008] A. Aiuppa, G. Giudice, S. Gurrieri, M. Luizzo, M. Burton, T. Caltabiano, A. J. S. McGonigle, G. Salerno, H. Shinohara, and M. Valenza, Total volatile flux from Mount Etna, Geophysical Research Letters, 35, L24302, 2008

[Grattan et al 2003] J. Grattan, M. Durand and S. Taylor, Illnes and elevated human mortality in Europe coincident with the Laki Fissure eruption. Geological Society, London, Special Publications 2003, **213**, pp 401-414, 2003

[Oman et al 2006] L. Oman, A. Robock, G. L. Stenchikov, Th. Thordarson, D. Koch, D. T. Shindell, and T. Gao, Modeling the distribution of the volcanic aerosol cloud from the 1783–1784 Laki eruption, Journal of Geophysical Research Vol. 111, 2006

[Stevenson et al 2003] D. S. Stevenson, C. E. Johnson, E. J. Highwood, V. Gauci, W. J. Collins, and R. G. Derwent, Atmospheric impact of the 1783–1784 Laki eruption: Part I Chemistry modelling, Atmos. Chem. Phys., 3, pp 487–507, 2003

[Ilmanlaatu] Air quality in Finland, $SO_2$-measurements for the Muonio-station, http://www.ilmanlaatu.fi/ilmanyt/nyt/ilmanyt.php?as=Suomi&rs=Valitse+kunta&ss=356&p=sulphurdioxide&pv=30.09.2014&h=14&et=graph&j=720&tj=3600&sc=500&ls=englanti, retrieved 7 March 2015.

[Stohl et al ] Stohl, A., Prata, A. J., Eckhardt, S., Clarisse, L., Durant, A., Henne, S., Kristiansen, N. I., Minikin, A., Schumann, U., Seibert, P., Stebel, K., Thomas, H. E., Thorsteinsson, T., Tørseth, K., and Weinzierl, B.: Determination of time- and height-resolved volcanic ash emissions and their use for quantitative ash dispersion modeling: the 2010 Eyjafjallajökull eruption, Atmos. Chem. Phys., 11, 4333–4351, doi:10.5194/acp-11-4333-2011, 2011.